\documentclass[10pt]{iopart}
\usepackage{graphicx}
\usepackage{xspace}				

\newcommand{\esa}{\textsl{ESA}\xspace}
\newcommand{\ieec}{\textsl{IEEC}\xspace}
\newcommand{\lisa}{\textsl{LISA}\xspace}
\newcommand{\lpf}{\textsl{\mbox{LPF}}\xspace}
\newcommand{\ltp}{\textsl{LTP}\xspace}

\begin{document}


\jl{6}

\title[Theoretical foundations for\ldots]{Theoretical foundations for on-ground
tests of \textsl{LISA PathFinder} thermal diagnostics}

\author{A Lobo$^{1,2}$\footnote[3]{To whom correspondence should be
addressed.}, M Nofrarias$^2$, J Ramos-Castro$^3$ and J Sanju\'an$^2$}

\address{$^1$ Institut de Ci\`encies de l'Espai, {\sl CSIC}}

\address{$^2$ Institut d'Estudis Espacials de Catalunya ({\sl IEEC\/}), Edifici
	{\sl Nexus}, Gran Capit\`a~2--4, 08034 Barcelona, Spain}

\address{$^3$ Departament d'Enginyeria Electr\`onica, {\sl UPC},
	Campus Nord, Edif.\ C4, Jordi Girona 1--3, 08034 Barcelona, Spain
	\ead{lobo@ieec.fcr.es}}

\date{\today}

\begin{abstract}
This paper reports on the methods and results of a theoretical analysis
to design an insulator which must provide a thermally quiet environment
to test on ground delicate temperature sensors and associated electronics.
These will fly on board \esa's \lisa Pathfinder (\lpf) mission as part of
the thermal diagnostics subsystem of the \lisa Test-flight Package (\ltp).
We evaluate the heat transfer function (in frequency domain) of a central
body of good thermal conductivity surrounded by a layer of a very poorly
conducting substrate. This is applied to assess the materials and
dimensions necessary to meet temperature stability requirements in the
metal core, where sensors will be implanted for test. The analysis is
extended to evaluate the losses caused by heat leakage through connecting
wires, linking the sensors with the electronics in a box outside the
insulator. The results indicate that, in spite of the very demanding
stability conditions, a sphere of outer diameter of the order one metre
is sufficient.
\end{abstract}
%
\submitto{\CQG}
%

\section{Introduction
\label{sec.1}}

\lisa Pathfinder (\lpf) is an \esa mission, whose main objective is
to put to test critical parts of \lisa (Laser Interferometer Space
Antenna), the first space borne gravitational wave (GW) observatory
\cite{bender}. The science module on board \lpf is the \lisa Test-flight
Package (\ltp)~\cite{lpfall}, which basically consists in two test masses
in nominally perfect geodesic motion (free fall), and a laser metrology
system, which reports on \emph{residual deviations} of the test masses'
actual motion from the ideal free fall, to a given level of accuracy
\cite{gerhar}.

In order to ensure that the test masses are not deviated from their
geodesic trajectories by external (non-gravitational) agents, a so
called Gravitational Reference System (GRS) is used~\cite{rita}. This
consists in position sensors for the masses which send signals to a set
of micro-thrusters; the latter take care of correcting as necessary the
spacecraft trajectory, so that at least one of the test masses remains
centred relative to the spacecraft at all times. The combination of the
GRS plus the actuators is known as \emph{drag-free} subsystem\footnote{
The term \emph{drag-free} dates back to the early days of space
navigation, when it was used to name a trajectory correction system
designed to compensate for the effect of atmospheric drag on satellites
in low altitude orbits.}.

The \emph{drag-free} is of course a central component of \lisa, and
needs to be operated at extremely demanding levels of accuracy. The
laser metrology system should then be sufficiently precise to measure
relative test mass deviations. The overall level of noise acceptable
for \lisa is defined in terms of rms acceleration spectral density,
and has been set to
\begin{equation}
 S_{a,{\rm LISA}}^{1/2}(\omega)\leq 3\!\times\!10^{-15}\,\left[
 1 + \left(\frac{\omega/2\pi}{3\ {\rm mHz}}\right)^{\!\!2}\right]\,
 {\rm m}\,{\rm s}^{-2}/\sqrt{\rm Hz} 
 \label{eq.1}
\end{equation}
in the frequency range
$\quad 10^{-4}\,{\rm Hz}\leq\omega/2\pi\leq 10^{-3}\,{\rm Hz}$.
This is equivalent to $S_h^{1/2}$\,$\sim$\,4$\times$10$^{-21}$
Hz$^{-1/2}$, with the same frequency dependence.

Because \lpf is a \emph{technological mission}, aimed to assess the
feasibility of \lisa, its ultimate goal has been relaxed to~\cite{toplev}
\begin{equation}
 S_{a,{\rm LPF}}^{1/2}(\omega)\leq 3\!\times\!10^{-14}\,\left[
 1 + \left(\frac{\omega/2\pi}{3\ {\rm mHz}}\right)^{\!\!2}\right]\,
 {\rm m}\,{\rm s}^{-2}/\sqrt{\rm Hz}
 \label{eq.2}
\end{equation}
in the frequency range $1\,{\rm mHz}\leq\omega/2\pi\leq 30\,{\rm mHz}$,
i.e., one order of magnitude less demanding, both in noise amplitude and
in frequency band.

Equation (\ref{eq.2}) gives the \emph{global} noise budget. This is
naturally made up of contributions from different perturbative agents,
such as temperature and magnetic field fluctuations, GRS and interferometer
noise, etc. As a general rule, a requirement on the magnitude of each of
the various perturbing factors is set at a 10\,\% fraction of the total.
In the case of temperature fluctuations, this is equivalent to
\begin{equation}
 S_{T}^{1/2}(\omega)\leq 10^{-4}\,{\rm K}/\sqrt{\rm Hz}\ ,
 \quad 1\,{\rm mHz}\leq \omega/2\pi \leq 30\,{\rm mHz}
 \label{eq.3}
\end{equation}

Because temperature stability is important, a decision has been taken to
place high precision thermometers in several strategic spots across the
\ltp ---as part of what is called \emph{Diagnostics Subsystem}~\cite{lobo}
\footnote{
The Diagnostics Subsystem of the \ltp also includes magnetometric
measurements and a charged particle flux detector.}. Such high precision
temperature measurements will be useful to identify the fraction of the
total system noise which is due to thermal fluctuations only, and this
will in turn provide important debugging information to assess the
performance of the \ltp.

\subsection{Temperature measurements
\label{sec.1-1}}

If the temperature gauges are to be sensitive to fluctuations at the
level given by (\ref{eq.3}) then clearly the entire measuring device
should be less noisy, typically by a factor of~10. This means that
such device, which includes both the sensors \emph{and} the associated
electronics, can generate a maximum level of noise of
\begin{equation}
 S_{T, {\rm sensor}}^{1/2}(\omega)\leq 10^{-5}\,{\rm K}/\sqrt{\rm Hz}\ ,
 \quad 1\,{\rm mHz}\leq \omega/2\pi \leq 30\,{\rm mHz}
 \label{eq.4}
\end{equation}

Research work is currently being conducted at \ieec (Barcelona, Spain)
to identify the appropriate sensors and design the better suited front
end electronics. But the prototype system needs of course to be tested
for compliance with equation~(\ref{eq.4}). Thus, in order to do a
meaningful test, the system must be sufficiently thermally isolated
that the observed fluctuations in the readout data can be attributed
\emph{solely} to sensor noise, rather than to a combination of it with
real ambient temperature fluctuations. This means temperature fluctiations
in the thermomters' placements should again be at least one order of
magnitude below the target sensitivity, equation~(\ref{eq.4}), or
\begin{equation}
 S_{T, {\rm testbed}}^{1/2}(\omega)\leq 10^{-6}\,{\rm K}/\sqrt{\rm Hz}\ ,
 \quad 1\,{\rm mHz}\leq \omega/2\pi \leq 30\,{\rm mHz}
 \label{eq.5}
\end{equation}

It turns out that 10$^{-6}\,{\rm K}/\sqrt{\rm Hz}$ is a truly demanding
temperature stability, orders of magnitude beyond the capabilities of
normal thermally regulated rooms. We thus need to design a specific
thermal insulator to shield the sensors from ambient temperature
fluctuations during the test process.

In the ensuing pages we describe in detail the insulator design. It is
extremely important to stress at this point that the performance of the
insulator, i.e., its ability to screen out ambient temperature fluctuations,
\emph{cannot be checked} experimentally, at least under working thermal
conditions in the laboratory. This is because, by definition, the insulator
is the tool to check the sensing instruments, \emph{not viceversa}: we need
to rely on the results of theoretical argumentation to make a decission on
which is the appropriate thermal insulator for our purposes. An experimental
verification of the model is only thinkable under much more extreme
conditions, where external temperature fluctuations are orders of magnitude
higher than the ones which will be met during the test.

\section{Thermal insulator design concept
\label{sec.2}}

\begin{figure}[b]
\centering
\includegraphics[width=6.1cm]{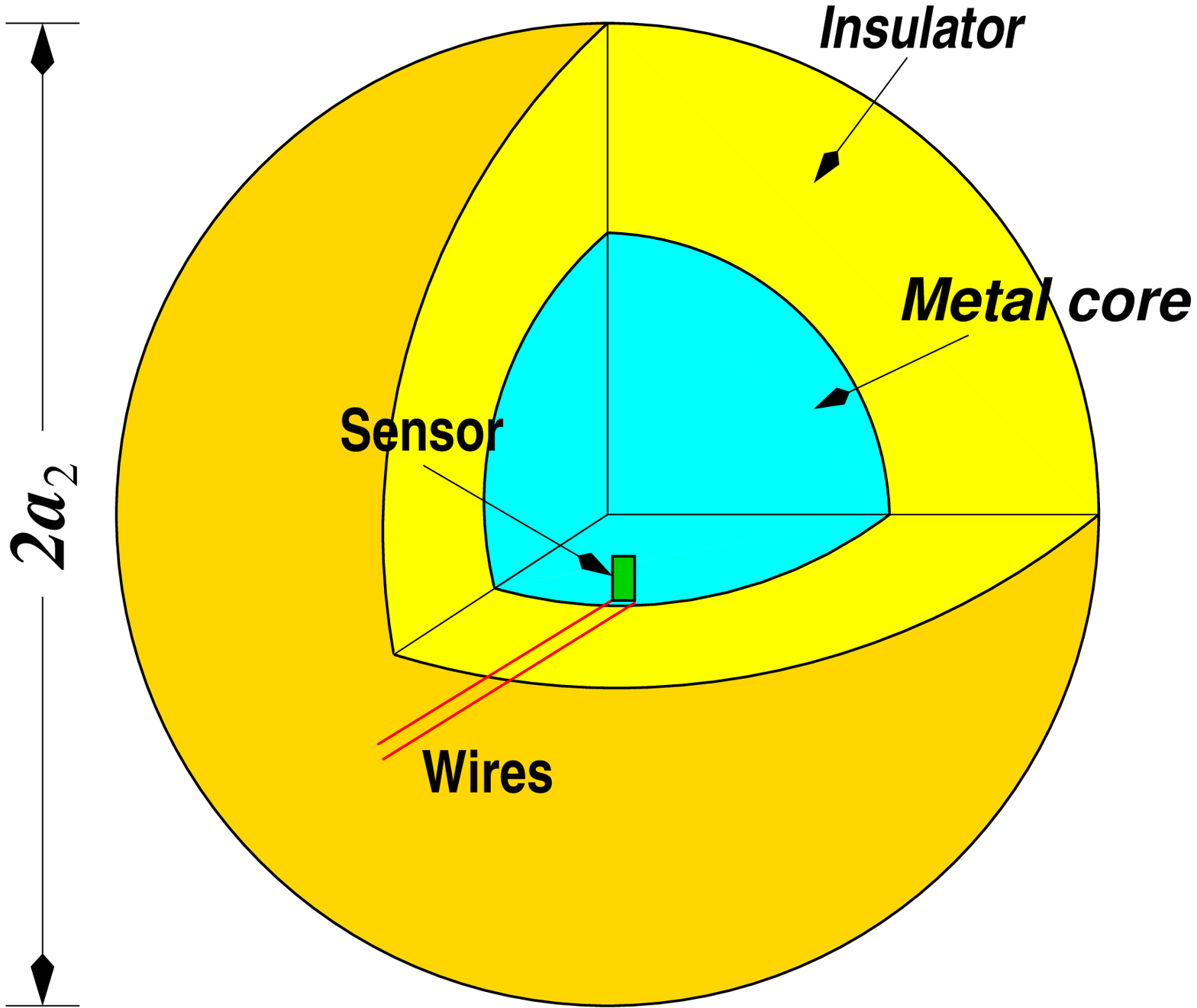}
\qquad
\includegraphics[width=4.9cm]{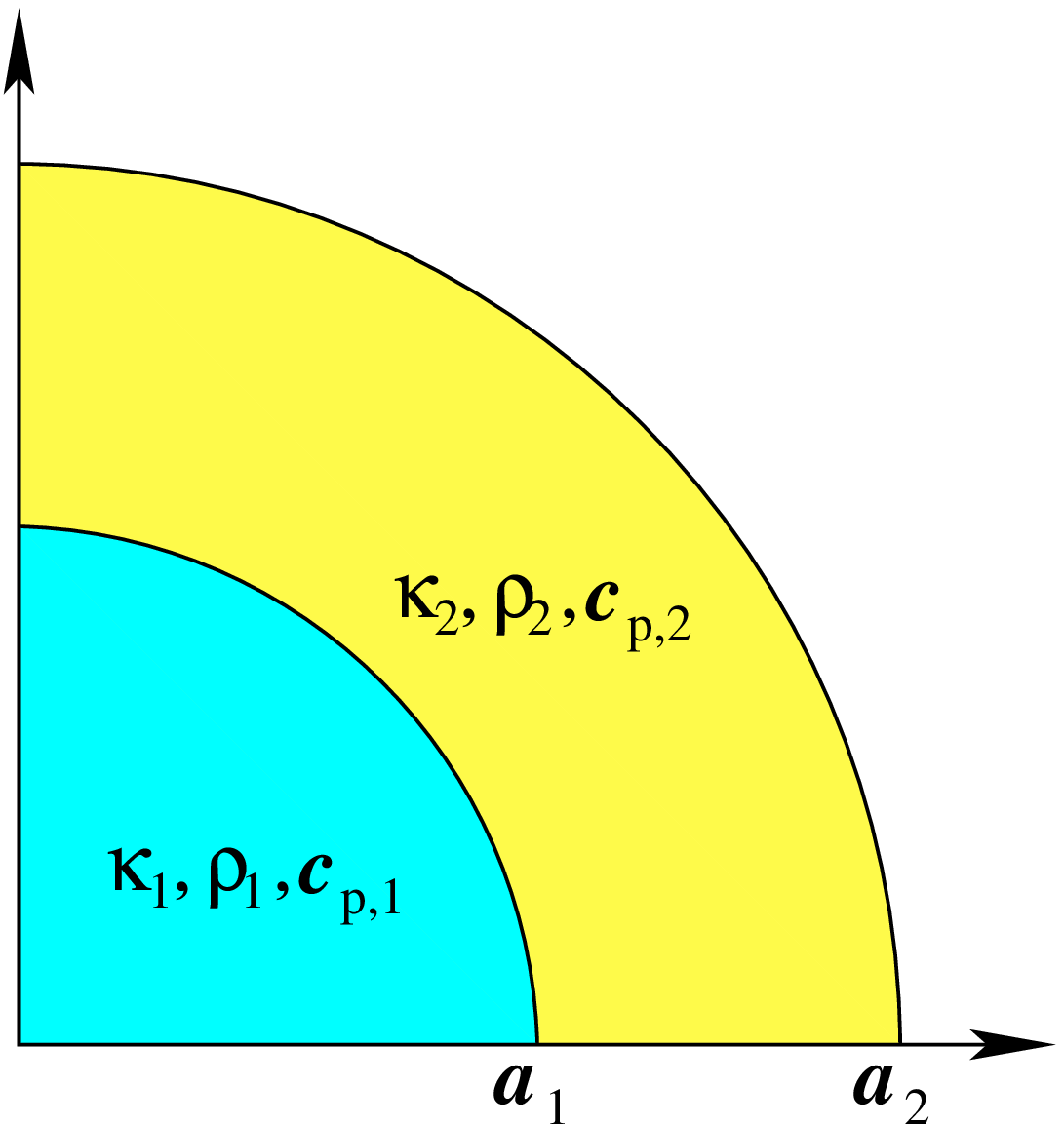}
\caption{Thermal insulator design concept. Left: 3D diagramme, including
sensor placement principle. Right: cut view, with notation convention
dictionary. \label{fig.1}}
\end{figure}

The idea of the insulator design is displayed in figure~\ref{fig.1}:
an interior metal core of good thermal conductivity is surrounded by
a thick layer of a poorly conductive material. The inner block ensures
thermal stability of the sensors attached to it, while the surrounding
substrate efficiently shields it from external temperature fluctuations
in the laboratory ambient. We propose a spherical shape for the sake
of simplicity of the mathematical analysis, even though this will be
eventually changed to cubic in the actual experimental device due to
practical feasibility issues.

\subsection{Mathematical model
\label{sec.2.1}}

The basic assumption of the mathematical analysis we shall present is
that heat flows from the interior of the insulator to the air outside,
and from the latter to the interior of the insulator, only by thermal
\emph{conduction}. This is a very realistic hypothesis in the context
of the experiment, as radiation mechanisms are certainly negligible
and convection should not play any significant role, either, since
the entire body is solid, and temperature fluctuations will be small
at all times anyway.

Let then $T({\bf x},t)$ be the temperature at time $t\/$ of a point
positioned at vector {\bf x} relative to the centre of the sphere.
$T({\bf x},t)$ thus satisfies Fourier's partial differential equation
\cite{carslaw}
\begin{equation}
 \rho c_{\rm p}\,\frac{\partial}{\partial t} T({\bf x},t) =
 \nabla\cdot\left[\kappa\nabla T({\bf x},t)\right]
 \label{eq.6}
\end{equation}
where $\rho$, $c_{\rm p}$ and $\kappa$ are the density, specific heat
and thermal conductivity, respectively, of the substrate. We shall
assume these are uniform values within each of the two materials making
up the insulating body, with abrupt changes in the interface. We can
thus represent them as discontinuous functions of the radial coordinate,
as follows:
\begin{equation}
 \rho, c_{\rm p}, \kappa({\bf x}) = \left\{\begin{array}{ll}
 \rho_1, c_{{\rm p}1}, \kappa_1 \quad & {\rm if}\ \ 0\leq r < a_1 \\
 \rho_2, c_{{\rm p}2}, \kappa_2 \quad & {\rm if}\ \ a_1\leq r < a_2
 \end{array}\right.
 \label{eq.7}
\end{equation}
with $r\/$\,$\equiv$\,$|{\bf x}|$. Initial and boundary conditions are
the following:
\begin{equation}
 T({\bf x},t=0) = 0\ ,\quad T(r=a_2,t) = T_0(\theta,\varphi;t)
 \label{eq.8}
\end{equation}
where $\theta\/$ and $\varphi\/$ are spherical angles which define positions
on the sphere's surface. The boundary temperature can be expediently
expressed as a multipole expansion:
\begin{equation}
 T_0(\theta,\varphi;t) = \sum_{lm}\,b_{lm}(t)\,Y_{lm}(\theta,\varphi)
 \label{eq.9}
\end{equation}
where $Y_{lm}(\theta,\varphi)$ are spherical harmonics, and $b_{lm}(t)$
are boundary multipole temperature components.

In practice, the boundary temperature will be \emph{randomly fluctuating},
therefore $b_{lm}(t)$ will be considered \emph{stochastic} functions of
time. We shall also reasonably assume them to be \emph{stationary Gaussian}
noise processes with known spectral densities, $S_{lm}(\omega)$.

As shown in the appendix, the frequency analysis of this problem leads to
a \emph{transfer function} expression of the temperature inside the body:
\begin{equation}
 \tilde T({\bf x},\omega) =
 \sum_{lm}\,H_{lm}({\bf x},\omega)\,\tilde b_{lm}(\omega)
 \label{eq.10}
\end{equation}
where \emph{tildes} (\,$\tilde{}$\,) stand for Fourier transforms, e.g.,
\begin{equation}
 \tilde T({\bf x},\omega)\equiv\int_{-\infty}^\infty\,
 T({\bf x},t)\,e^{-i\omega t}\,dt
 \label{eq.11}
\end{equation}
etc. If we make the further assumption that different multipole temperature
fluctuations at the boundary are \emph{uncorrelated}, i.e.,
\begin{equation}
 \langle\tilde b^*_{l'm'}(\omega)\,\tilde b_{lm}(\omega)\rangle
 = S_{lm}(\omega)\,\delta_{l'l}\,\delta_{m'm}
 \label{eq.12b}
\end{equation}
then the spectral density of fluctuations at any given point inside the
insulating body is given by
\begin{equation}
 S_T({\bf x},\omega) =
 \sum_{lm}\,\left|H_{lm}({\bf x},\omega)\right|^2\,
 S_{lm}(\omega)
 \label{eq.12}
\end{equation}

It is ultimately the spectral density $S_T({\bf x},\omega)$ which has to
comply with the requirement expressed by equation~(\ref{eq.4}). Based on
knowledge (by direct measurement) of ambient laboratory temperature
fluctuations, equation~(\ref{eq.12}) will provide the guidelines, as
regards materials and dimensions, for the actual design of a suitable
insulator jig.

\section{Homogeneous boundary conditions
\label{sec.3}}

Thermal conditions in the laboratory are rather \emph{homogeneous}. This
means that the boundary temperature fluctuations will be in practice
essentially independent of the angles $\theta\/$ and $\varphi$, i.e.,
\begin{equation}
 T_0(\theta,\varphi;t) = B(t)
 \label{eq.13}
\end{equation}
and consequently the generic expansion equation~(\ref{eq.9}) includes
only the \emph{monopole} term, hence
\begin{equation}
 b_{00}(t) = \sqrt{4\pi}\,B(t)
 \label{eq.14}
\end{equation}

The temperature $T({\bf x},\omega)$ in this case will only depend on
radial depth, $r$, therefore,
\begin{equation}
 \tilde T(r,\omega) = H(r,\omega)\,\tilde B(\omega)
 \label{eq.15}
\end{equation}
with $H(r,\omega)$\,$\equiv$\,$\sqrt{4\pi}\,H_{00}({\bf x},\omega)$.
According to equation~(\ref{eq.a13}) of the Appendix, this is
\begin{equation}
 \hspace*{-0.6 cm}
 H(r,\omega) = \left\{\begin{array}{ll}
 \xi_0(\omega)\,j_0(\gamma_1r)\ , &
 0\leq r \leq a_1 \\[1 em]
 \eta_0(\omega)\,j_0(\gamma_2r) + 
 \zeta_0(\omega)\,y_0(\gamma_2r)\ , &
 a_1\leq r \leq a_2 \end{array}\right.
 \label{eq.16}
\end{equation}

\begin{figure}[b]
\centering
\includegraphics[width=7.8cm]{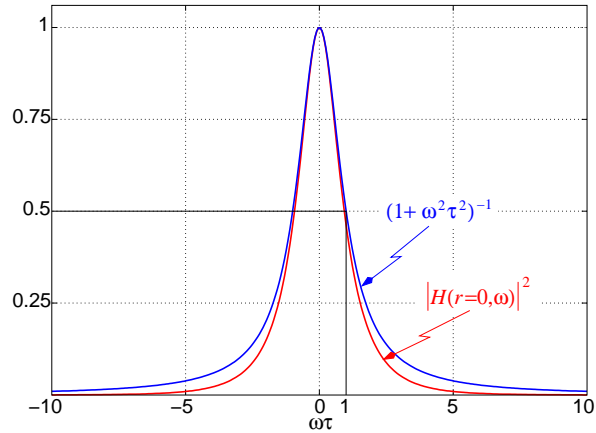}
\caption{Frequency response at the centre ($r\/$\,=\,0) of a spherical
thermal insulator ---as given in equation~(\protect\ref{eq.16})---, along
with the frequency response of a low-pass filter of order one and same
frequency cut-off, $\omega_{\rm cut-off}$\,=\,$1/\tau$. \label{fig.2}}
\end{figure}

This is a low-pass filter transfer function ---even though the cumbersome
frequency dependencies involved in the expressions above do not make it
immediately obvious. A plot of the square modulus of $H(r,\omega)$ is
shown in figure~\ref{fig.2} for $r\/$\,=\,0 (red curve). The figure also
shows a low-pass filter of the first order with the same frequency cut-off,
$|H_{\rm 1st\ order}(\omega)|^2$\,=\,$(1+\omega^2\tau^2)^{-1}$, for conceptual
comparison (blue curve).

The most salient feature emerging out of the plot is the stronger drop
in $H(r,\omega)$ at the high frequency tails. The latter can be easily
assessed in quantitative detail, and the result is
%
%
%
%
%
%
\begin{equation}
 |H(0,\omega)|\sim\omega\tau\,e^{-\sqrt{\omega\tau}}
 \label{eq.19}
\end{equation}
where $\tau$ is the filter's time constant ---a complicated function of
the insulator's physical and geometric properties, to be discussed below.

As already mentioned in the Introduction section, to test the temperature
sensors and electronics we need a very strong noise suppression factor in
the \ltp frequency band. A look at figure~\ref{fig.2} readily shows that
high damping factors require such frequency band to lie in the filter's
tails. The thermal insulator should therefore be designed in such a way
that its time constant $\tau\/$ be sufficiently large to ensure that the
\ltp frequencies are high enough compared to 1/$\tau\/$. The exponential
drop in the transfer function shown by equation~(\ref{eq.19}) makes the
filter actually feasible with reasonable dimensions.

\section{Numerical analysis
\label{sec.4}}

In this section we consider the application of the above formalism to
obtain practically useful numbers for the actual implementation of a
real insulator device which complies with the needs of our experiment.

First of all, a selection of an \emph{aluminum} core surrounded by a
layer of \emph{polyurethane} was made. Aluminum is a good heat conductor
and is easy to work with in the laboratory; polyurethane is a good
insulator and is also convenient to handle, as it can be foamed to any
desired shape from canned liquid. Other alternatives are certainly
possible, but this appears sufficiently good and we shall therefore
only make reference to this specific one.

The relevant physical properties of aluminium and polyurethane are
specified in table~\ref{tab.1}.

\begin{table}[h!]
\begin{center}
\begin{tabular}{lccc}
& Density
& Specific heat
& Thermal conductivity \\
& $\rho$ (kg\,m$^{-3}$)
& $c_{\rm p}$ (J\,kg$^{-1}$\,K$^{-1}$)
& $\kappa$ (W\,m$^{-1}$\,K$^{-1}$) \\[1ex]
\hline \\[-1.3ex]
{\sf Aluminum} & 2700 & 900 & 250 \\
{\sf Polyurethane} & 35 & 1000 & 0.04
\end{tabular}
\caption{Density, specific heat and thermal conductivity of aluminium
and polyurethane. Units are given in the International System.
\label{tab.1}}
\end{center}
\end{table}

\begin{figure}
\centering
\includegraphics[width=9.7cm]{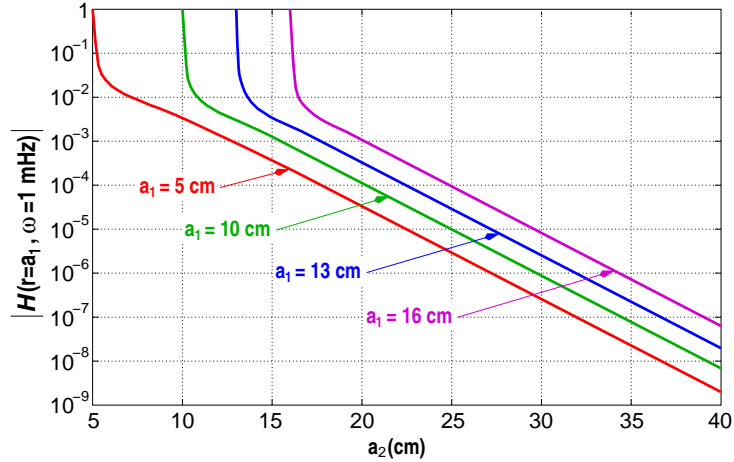}
\caption{Amplitude damping factor on the interface between aluminium
and polyurethane at 1 mHz, for various sizes of the insulating device.
\label{fig.3}}
\end{figure}

Figure \ref{fig.3} plots the \emph{amplitude damping coefficient} of
the insulator block, $|H(r,\omega)|$, at the lower end of the \ltp
frequency band, i.e., 1 mHz, and at the interface position,
$r\/$\,=\,$a_1$. Each of the curves corresponds to a fixed value of
the latter, and is represented as a function of the outer radius of
the insulator. This choice is useful because the sensors are implanted
for test on the surface of the aluminium core, and also because at
higher frequencies thermal damping is stronger. So in practice the
actual damping power of the device will be the one plotted, and better
at the higher frequencies in the measuring bandwidth. The figure clearly
shows that the assymptotic regime of equation~(\ref{eq.19}) is quite
early established.

The choice of dimensions for the insulating body must of course ensure
that the minimum requirement, equation~(\ref{eq.5}) is met. For this,
a primary consideration is the size of the ambient temperature
fluctuations in the site where the experiment is made. Dedicated
measurements in our laboratory showed that
\begin{equation}
 S_{T, {\rm ambient}}^{1/2}(\omega)\sim 10^{-1}\,{\rm K}/\sqrt{\rm Hz}\ ,
 \quad 1\,{\rm mHz}\leq \omega/2\pi \leq 30\,{\rm mHz}
 \label{eq.20}
\end{equation}

We therefore need to implement a device such that
$|H(a_1,\omega)|$\,$\leq$\,10$^{-5}$ throughout the measuring bandwidth
(MBW). Suitable dimensions can then be readily read off figure~\ref{fig.3},
and various alternatives are possible, as seen. Before making a decission,
however, we need to make an additional estimate of the heat leakage
down the electric wires which connect the temperature sensors with
the elctronics, which lies of course outside the insulator. We come
to this next.

\subsection{Heat leakage through connecting wires}

We use a simple model, consisting in assuming the connecting wires
behave as straight metallic rods which connect the central aluminum
core with the electronics, placed in the external laboratory ambient.
Because the polyurethane provides a very stable insulation, we can
neglect the lateral flux, hence only a unidimensional heat flow
needs to be considered. For this, the following equation relates
the heat flux to the temperature difference between the two wires'
edges:
\begin{equation}
 \dot{Q}(t) = \kappa_{\rm wire}\,\frac{\pi R_{\rm wire}^2}{\ell_{\rm wire}}\;
 [T(a_2,t)-T(a_1,t)]
 \label{eq.21}
\end{equation}
where $\kappa_{\rm wire}$ is the thermal conductivity of the wire,
$R_{\rm wire}$ its transverse radius, and $\ell_{\rm wire}$ its length
\emph{inside} the polyurethane layer.

On the other hand, the heat flux results in temperature variations in
the metal core, given by
\begin{equation}
 \dot{Q}(t) = \rho_1c_{{\rm p}1}V_1\,
 \frac{\partial T}{\partial t}(a_1,t)
 \label{eq.22}
\end{equation}
where $V_1$\,=\,$4\pi a_1^3/3$ is the volume of the metal core. Equating
the above expressions we find
\begin{equation}
 \kappa_{\rm wire}\,\frac{\pi R_{\rm wire}^2}{\ell_{\rm wire}}\;
 [T(a_2,t)-T(a_1,t)] =
 \rho_1c_{{\rm p}1}V_1\, \frac{\partial T}{\partial t}(a_1,t)
 \label{eq.23}
\end{equation}

For fluctuating temperatures, we can now obtain the relationship between
the spectral density at the aluminium core and the ambient, due to heat
conduction along the wire:
\begin{equation}
 S_{T,{\rm wire}}^{1/2}(a_1,\omega) = |H_{\rm wire}(\omega)|\,
 S_{T,{\rm ambient}}^{1/2}(\omega)
 \label{eq.24}
\end{equation}
where
\begin{equation}
 |H_{\rm wire}(\omega)|\simeq\frac{\pi}{\omega}\,
 \frac{\kappa_{\rm wire}\,R_{\rm wire}^2}
 {\rho_1c_{{\rm p}1}V_1\,\ell_{\rm wire}}
 \label{eq.25}
\end{equation}
and where the approximation has been made that the temperature fluctuations
at the inner end of the wire are much smaller than those at the outer
end, due to the presence of the polyurethane layer.

In practice, there will be several sensors for test inside the insulator.
Under the hypothesis made that no lateral heat flux is relevant, the transfer
function for a bundle of $N\/$ of wires is, at most, $N\/$ times that of a
single wire. Thus,
\begin{equation}
 |H_{N{\rm wires}}(\omega)| = \frac{3N}{\omega/2\pi}\,
 \frac{\kappa_{\rm wire}\,R_{\rm wire}^2}
 {8\pi\rho_1c_{{\rm p}1}a_1^3\,\ell_{\rm wire}}
 \label{eq.26}
\end{equation}

Let us consider numerical values in this expression. We use thin
copper wires ($\kappa_{\rm Cu}$\,=\,401\,Wm$^{-1}$K$^{-1}$) of
radius $R_{\rm wire}$\,=\,0.1\,mm, and assume some fiducial
parameters for the size of the aluminium core, $a_1$, the wire
length, $\ell_{\rm wire}$, the number of connecting wires, $N$,
and the frequency, $\omega/2\pi$. The following obtains:
\begin{equation}
 \hspace*{-1.2 cm}
 |H_{N{\rm wires}}(\omega)| = 1.1\times 10^{-5}\,
 \left(\frac{N}{30}\right)\!
 \left(\frac{a_1}{13\ {\rm cm}}\right)^{\!-3}\!
 \left(\frac{\ell_{\rm wire}}{25\ {\rm cm}}\right)^{\!-1}\!
 \left(\frac{\omega/2\pi}{1\ {\rm mHz}}\right)^{\!-1}
 \label{eq.27}
\end{equation}


This result indicates that, for laboratory fluctuations in the level of
equation~(\ref{eq.20}), leakage through wiring causes fluctuations in
the temperature sensors of about 10$^{-6}$\,K/$\sqrt{\rm Hz}$,
equation~\eref{eq.24}, which is compliant with the requirement of
stability of equation~\eref{eq.5}. The most sensitive parameter in
the above expression is the size of the metal core, and this determines
the need to make it somewhat large. The length of the wires has been
taken to be 25~cm, but this does not necessarily mean we need
$a_2$\,=\,38~cm (assuming the radius of the aluminum core is
$a_1$\,=\,13~cm), because the wires can be partly wound inside the
polyurethane layer to further protect the system against leakage.
In fact, this wire lengthening is an easy way to improve attenuation.

As regards frequency dependence, compliance is guaranteed in the entire
MBW if it is at its lower end: indeed, not only $|H_{\rm wire}(\omega)|$
decreases as $\omega^{-1}$, also ambient noise fluctuations drop below
10$^{-1}$\,K/$\sqrt{\rm Hz}$ at higher frequencies.

\section{Conclusions}

Temperature fluctuation measurement is very demanding in the \ltp, and
subsequently \lisa, as reflected by equation~\eref{eq.4}. Accordingly,
very delicate sensor and associated electronics must be designed, and
of course tested in ground before boarding.

However, even the best laboratory conditions are orders of magnitude
worse than the above requirement, so meaningful tests of the temperature
sensing system cannot be tested without suitably screening the sensors
from ambient temperature fluctuations. We have addressed how this can
be accomplished by means of an insulating system consisting of a central
metallic core surrounded by a thick layer of a very poorly conducting
material. The latter provides good thermal insulation, while the central
core, having a large thermal inertia, ensures stability of the sensors'
environemnt. The choice of materials is flexible, so aluminium and
polyurethane, which are easily available in the market, has been
adopted. Thereafter, the dimensions need to be fixed.

The appropriate sensors for the needs are temperature sensitive resistors,
more specifically thermistors ---also known as NTCs. It appears that,
because these sensors need to be wired to external electronics, heat leakage
through such wires is an effect which needs to be quantitatively assessed
to prevent losses. We have analysed this problem, and concluded that it
strongly depends on the central metallic core size, and imposes that it
be somewhat large.

Laboratory ambient temperature fluctuations, determined by dedicated
\emph{in situ} measurements, are of the order of 10$^{-1}$\,K/$\sqrt{\rm Hz}$
at 1~mHz, and dropping at higher frequencies within the MBW. The required
stability conditions at the sensors, attached at the core's surface, thus
need an attenuation factor of 10$^{-5}$, or better. Our analysis determines
that a central aluminium core of 13~cm of radius, surrounded by a concentric
layer of polyurethane 15--20~cm thick, comfortably provides the needed
thermal screening which guarantees a meaningful test of the sensors'
performance.

The results of this paper are based on modelling. Because our aim is to
produce a very stable thermal environment for the temperature sensors,
we cannot check \emph{experimentally} the correctness of our conclusions.
We must instead rely on the validity of the hypotheses made ---essentially
that heat only flows by thermal conduction--- and on the underlying physical
laws which govern heat conduction. Even though there is good reason to
believe that both are sufficiently accurate, unexpected behaviour e.g. at
the interface between the metal core and the insulator, may partly distort
the results. Direct measurements with very large temperature gradients
applied across the insulating device are envisaged, and will be reported
elsewhere as an auxiliary independent test of the model.

\ack

We want to thank Albert Tom\`as, from {\sl NTE}, for discussions on
the subject of this paper. Support for this work came from Project
ESP2004-01647 of Plan Nacional del Espacio of the Spanish Ministry
of Education and Science (MEC). MN acknowledges a grant from Generalitat
de Catalunya, and JS a grant from MEC.

\appendix

\section{Thermal insulator frequency response functions
\label{sec.a1}}

Here we present some mathematical details of the solution to the Fourier
problem, equations~(\ref{eq.6})-(\ref{eq.9}). We first of all Fourier
transform equations~(\ref{eq.6}) and (\ref{eq.9}):
\begin{equation}
 i\omega\,\rho c_{\rm p}\,\tilde T({\bf x},\omega) =
 \nabla\cdot\left[\kappa\nabla\tilde T({\bf x},\omega)\right]
 \label{eq.a1}
\end{equation}

\begin{equation}
 \tilde T_0(\theta,\varphi;\omega) =
 \sum_{l=0}^\infty\sum_{m=-l}^l\,\tilde b_{lm}(\omega)\,Y_{lm}(\theta,\varphi)
 \label{eq.a2}
\end{equation}

Equation (\ref{eq.a1}) can be recast in the form
\begin{equation} 
 \left(\nabla^2 + \gamma_1^2\right)\,\tilde T({\bf x},\omega) = 0\ ,
 \quad 0\leq r\leq a_1
 \label{eq.a3a}
\end{equation}

\begin{equation} 
 \left(\nabla^2 + \gamma_2^2\right)\,\tilde T({\bf x},\omega) = 0\ ,
 \quad a_1\leq r\leq a_2
 \label{eq.a3b}
\end{equation}
where $r\/$\,$\equiv$\,$|{\bf x}|$, and
\begin{equation}
 \gamma_1^2\equiv -i\omega\,\frac{\rho_1 c_{\rm p,1}}{\kappa_1}\ ,\quad
 \gamma_2^2\equiv -i\omega\,\frac{\rho_2 c_{\rm p,2}}{\kappa_2}
 \label{eq.a4}
\end{equation}

To these, matching conditions at the interface\footnote{
The temperature and the \emph{heat flux} should be continuous across
the interface.}
and boundary conditions must be added:
\begin{equation}
 \tilde T(r=a_1-0,\omega) = \tilde T(r=a_1+0,\omega)
 \label{eq.a7a}
\end{equation}

\begin{equation}
 \kappa_1\,\frac{\partial \tilde T}{\partial r}(r=a_1-0,\omega) =
 \kappa_2\,\frac{\partial \tilde T}{\partial r}(r=a_1+0,\omega)
 \label{eq.a7b}
\end{equation}

\begin{equation}
 \tilde T(r=a_2,\omega) = \tilde T_0(\theta,\varphi;\omega)
 \label{eq.a7c}
\end{equation}

Equations (\ref{eq.a3a}) and (\ref{eq.a3b}) are of the Helmholtz kind. Their
solutions are thus respectively given by
\begin{equation}
 \hspace*{-2.25 cm}
 \tilde T({\bf x},\omega) = \left\{\begin{array}{ll}
 \displaystyle
 \sum_{lm}\,A_{lm}(\omega)\,j_l(\gamma_1r)\,Y_{lm}(\theta,\varphi)\ , &
 0\leq r \leq a_1 \\[1.7 em]
 \displaystyle
 \sum_{lm}\,\left[C_{lm}(\omega)\,j_l(\gamma_2r) + 
                  D_{lm}(\omega)\,y_l(\gamma_2r)\,\right]\,
                  Y_{lm}(\theta,\varphi)\ , &
 a_1\leq r \leq a_2 \end{array}\right.
 \label{eq.a5}
\end{equation}
where $j_l\/$ and $y_l\/$ are spherical Bessel functions \cite{as72},
\begin{equation}
 \hspace*{-0.8 cm}
 j_l(z) = z^l\,\left(-\frac{1}{z}\,\frac{d}{dz}\right)^{\!\!l}\,
 \frac{\sin z}{z}\ ,\quad
 y_l(z) = -z^l\,\left(-\frac{1}{z}\,\frac{d}{dz}\right)^{\!\!l}\,
 \frac{\cos z}{z}
 \label{eq.a6}
\end{equation}
and the coefficients $A_{lm}(\omega)$, $C_{lm}(\omega)$ and $D_{lm}(\omega)$
are to be determined by equations~(\ref{eq.a7a})--(\ref{eq.a7c}). These can
be expanded as follows, respectively:
\begin{eqnarray}
 \sum_{lm}\,A_{lm}(\omega)\,j_l(\gamma_1a_1)\,Y_{lm}(\theta,\varphi)\ =
 & & \nonumber \\
 \ \ =\ \sum_{lm}\,\left[C_{lm}(\omega)\,j_l(\gamma_2a_1) +
                  D_{lm}(\omega)\,y_l(\gamma_2a_1)\,\right]\,
                  Y_{lm}(\theta,\varphi) & &
 \label{eq.a8a}
\end{eqnarray}

\begin{eqnarray}
 \kappa_1\gamma_1\,
 \sum_{lm}\,A_{lm}(\omega)\,j'_l(\gamma_1a_1)\,Y_{lm}(\theta,\varphi)\ =
 & & \nonumber \\
 \ \ =\ \kappa_2\gamma_2\,
 \sum_{lm}\,\left[C_{lm}(\omega)\,j'_l(\gamma_2a_1) +
                  D_{lm}(\omega)\,y'_l(\gamma_2a_1)\,\right]\,
                  Y_{lm}(\theta,\varphi)
 \label{eq.a8b}
\end{eqnarray}

\begin{eqnarray}
 \sum_{lm}\,\left[C_{lm}(\omega)\,j_l(\gamma_2a_2) + 
                  D_{lm}(\omega)\,y_l(\gamma_2a_2)\,\right]\,
                  Y_{lm}(\theta,\varphi)\ =
 & & \nonumber \\
 \ \ =\ \sum_{lm}\,\tilde b_{lm}(\omega)\,Y_{lm}(\theta,\varphi)
 \label{eq.a8c}
\end{eqnarray}

Because of the completeness property of the spherical harmonics, the
above equations completely determine the coefficients $A_{lm}(\omega)$,
$C_{lm}(\omega)$ and $D_{lm}(\omega)$. The result is
\begin{equation}
 \hspace*{-2 cm}
 A_{lm}(\omega) = \xi_l(\omega)\,\tilde b_{lm}(\omega)\ ,\ \ 
 C_{lm}(\omega) = \eta_l(\omega)\,\tilde b_{lm}(\omega)\ ,\ \ 
 D_{lm}(\omega) = \zeta_l(\omega)\,\tilde b_{lm}(\omega)
 \label{eq.a9}
\end{equation}
with
\begin{equation}
 \hspace*{-1 cm}
 \xi_l(\omega) = \frac{1}{\Delta_l(\omega)}\,\left[
 \kappa_2\gamma_2\,j_l(\gamma_2a_1)\,y'_l(\gamma_2a_1) -
 \kappa_2\gamma_2\,j'_l(\gamma_2a_1)\,y_l(\gamma_2a_1)\right]
 \label{eq.a9a}
\end{equation}

\begin{equation}
 \hspace*{-1 cm}
 \eta_l(\omega) = \frac{1}{\Delta_l(\omega)}\,\left[
 \kappa_2\gamma_2\,j_l(\gamma_1a_1)\,y'_l(\gamma_2a_1) -
 \kappa_1\gamma_1\,j'_l(\gamma_1a_1)\,y_l(\gamma_2a_1)\right]
 \label{eq.a9b}
\end{equation}

\begin{equation}
 \hspace*{-1. cm}
 \zeta_l(\omega) = \frac{1}{\Delta_l(\omega)}\,\left[
 \kappa_1\gamma_1\,j_l(\gamma_2a_1)\,j'_l(\gamma_1a_1) -
 \kappa_2\gamma_2\,j'_l(\gamma_2a_1)\,j_l(\gamma_1a_1)\right]
 \label{eq.a9c}
\end{equation}
and
\begin{eqnarray}
 \hspace*{-1 cm}
 \Delta_l(\omega) & = &
 \ \kappa_1\gamma_1\,j'_l(\gamma_1a_1)\,\left[
 j_l(\gamma_2a_1)\,y_l(\gamma_2a_2) -
 j_l(\gamma_2a_2)\,y_l(\gamma_2a_1)\right]\ + \nonumber \\
 & + &
 \ \kappa_2\gamma_2\,j_l(\gamma_1a_1)\,\left[
 j_l(\gamma_2a_2)\,y'_l(\gamma_2a_1) -
 j'_l(\gamma_2a_1)\,y_l(\gamma_2a_2)\right]
 \label{eq.a10}
\end{eqnarray}

When the above results, equations~(\ref{eq.a9a}) through (\ref{eq.a10}),
are inserted back into equation~(\ref{eq.a5}) the result stated in
equation~(\ref{eq.10}) in the main text obtains, i.e.,
\begin{equation}
 \tilde T({\bf x},\omega) =
 \sum_{lm}\,H_{lm}({\bf x},\omega)\,\tilde b_{lm}(\omega)
 \label{eq.a11}
\end{equation}
where
\begin{equation}
 \hspace*{-1.8 cm}
 H_{lm}({\bf x},\omega) = \left\{\begin{array}{ll}
 \xi_l(\omega)\,j_l(\gamma_1r)\,Y_{lm}(\theta,\varphi)\ , &
 0\leq r \leq a_1 \\[1 em]
 \left[\eta_l(\omega)\,j_l(\gamma_2r) + 
                  \zeta_l(\omega)\,y_l(\gamma_2r)\,\right]\,
                  Y_{lm}(\theta,\varphi)\ , &
 a_1\leq r \leq a_2 \end{array}\right.
 \label{eq.a12}
\end{equation}

For monopole only boundary conditions, equation~(\ref{eq.15}), the
transfer function is
\begin{equation}
 \hspace*{-0.6 cm}
 H(r,\omega) = \left\{\begin{array}{ll}
 \xi_0(\omega)\,j_0(\gamma_1r)\ , &
 0\leq r \leq a_1 \\[1 em]
 \eta_0(\omega)\,j_0(\gamma_2r) + 
                  \zeta_0(\omega)\,y_0(\gamma_2r)\ , &
 a_1\leq r \leq a_2 \end{array}\right.
 \label{eq.a13}
\end{equation}



\section*{References}

\begin{thebibliography}{99}

\bibitem{bender} Bender P \emph{et al} 2000 Laser Interferometer Space
	Antenna: A Cornerstone Mission for the observation of Gravitational
	Waves, ESA report no.\ ESA-SCI(2000)11

\bibitem{lpfall} Anza S \emph{et al} 2005 The \ltp Experiment on the
	\lisa Pathfinder Mission \CQG {\bf 22}, S125-38

\bibitem{gerhar} Heinzel G \emph{et al} 2004 The \ltp interferometer and
	phasemeter \CQG {\bf 21}, S581-88

\bibitem{rita} Dolesi R \emph{et al} 2003 Gravitational sensor for \lisa
	and its technology demonstration mission \CQG {\bf 20}, S99-108

\bibitem{toplev} Vitale S 2005 Science Requirements and Top-level Architecture
	Definition for the Lisa Technology Package (\ltp) on Board \lisa
	Pathfinder (SMART-2) \lpf report no.\ LTPA-UTN-ScRD-Iss003-Rev1

\bibitem{lobo} Lobo A 2005 DDS Science Requirements Document \lpf report
	no.\ S2-IEEC-RS-3002

\bibitem{carslaw} Carslaw HS and Jaeger JC 1986 \emph{Conduction of heat
	in solids} (Oxford University Press)

\bibitem{as72} Abramowitz M and Stegun IA 1972 \emph{Handbook of Mathematical
	Functions} (Dover, New York)

\end{thebibliography}
\end{document}